\documentclass[aps,prl,twocolumn,showpacs,superscriptaddress,nofootinbib,groupedaddress]{revtex4} 
\usepackage{dcolumn}   
\usepackage{pstricks}
\usepackage{color}
\usepackage{graphicx}
\usepackage{amsmath}
\usepackage{amssymb}
\usepackage[colorlinks=true,allcolors=darkred, pdfborder={0 0 0}]{hyperref}
\usepackage[T1]{fontenc}
\usepackage[latin9]{inputenc}
\usepackage{array}
\usepackage{booktabs}
\usepackage{mathrsfs}
\usepackage{multirow}
\usepackage{ulem}

\definecolor{darkred}{rgb}{0.6,0,0}

\definecolor{darkred}{rgb}{0.6471, 0.1098, 0.1882} 

\graphicspath{{figs/}}

\def\gsim{\raise0.3ex\hbox{$\;>$\kern-0.75em\raise-1.1ex\hbox{$\sim\;$}}}
\def\lsim{\raise0.3ex\hbox{$\;<$\kern-0.75em\raise-1.1ex\hbox{$\sim\;$}}}

\def\beqn#1{\begin{equation}\label{#1}}
\def\eeqn{\end{equation}}

\def\beqa#1{\begin{eqnarray}\label{#1}}
\def\eeqa{\end{eqnarray}}

\newcommand{\fig}[1]{Fig.~\ref{fig:#1}}
\newcommand{\tab}[1]{Table~\ref{tab:#1}}
\newcommand{\eq}[1]{Eq.~\ref{eq:#1}}

\def\Mo{M_8}

\def\vev#1{\left\langle #1\right\rangle}

\def\g{\gamma}
\def\g5{\gamma_5}
\def\21{SU(2) $\otimes$ U(1) }

\def\TrTrOne{ $SU(3)_C$ $\otimes$ $SU(3)_L$ $\otimes$ $U(1)_X$}
\def\TrOne{ $SU(3)_L$ $\otimes$ $U(1)_X$}
\def\TwoOne{ $SU(2)_L$ $\otimes$ $U(1)_Y$}

\def\Z2{$\mathcal{Z_2}$}

\def\vev#1{\left\langle #1\right\rangle}

\begin{document}

\title{Small neutrino masses and gauge coupling unification}


\author{Sofiane M. Boucenna}
\email{boucenna@ific.uv.es}
\affiliation{Instituto de F\'{\i}sica Corpuscular (CSIC-Universitat de Val\`{e}ncia), Apdo. 22085, E-46071 Valencia, Spain.}
\author{Renato M. Fonseca}
\email{renato.fonseca@ific.uv.es}
\affiliation{Instituto de F\'{\i}sica Corpuscular (CSIC-Universitat de Val\`{e}ncia), Apdo. 22085, E-46071 Valencia, Spain.}
\author{F\'elix Gonz\'alez-Canales}
\email{felix.gonzalez@ific.uv.es}
\affiliation{Instituto de F\'{\i}sica Corpuscular (CSIC-Universitat de Val\`{e}ncia), Apdo. 22085, E-46071 Valencia, Spain.}
\author{Jos\'e W.F. Valle}
\email{valle@ific.uv.es}
\affiliation{Instituto de F\'{\i}sica Corpuscular (CSIC-Universitat de Val\`{e}ncia), Apdo. 22085, E-46071 Valencia, Spain.}
\date{\today}


\begin{abstract}
\noindent

The physics responsible for gauge coupling unification may also induce
small neutrino masses. We propose a novel gauge mediated radiative
seesaw mechanism for calculable neutrino masses. These arise from
quantum corrections mediated by new \TrTrOne~ (3-3-1) gauge bosons and
the physics driving gauge coupling unification.  Gauge couplings unify
for a 3-3-1 scale in the TeV range, making the model directly testable
at the LHC.
\end{abstract}

\maketitle

\section*{Preliminaries}

The fact that gauge coupling unification is a ``near-miss'' within the
Standard Model (SM) provides an indication in favor of the idea of
unification~\cite{PhysRevLett.33.451}. Likewise, the existence of
neutrino masses, required to account for neutrino oscillation
data~\cite{Forero:2014bxa}, also provides another motivation towards
unified or GUT~(Grand Unified Theory)-like extensions of the
SM. However, the most characteristic feature of GUT-type unification,
namely matter instability, has so far defied experimental
confirmation~\citep{1674-1137-38-9-090001}. On the other hand, neither
the generation of neutrino masses nor the tilting in the evolution of
the gauge couplings require unification in the conventional
sense.
For instance, it is well known that the gauge couplings merge in the
minimal supersymmetric extension the SM, provided that supersymmetric
states lie around the TeV
scale~\cite{amaldi:1991cn,1674-1137-38-9-090001}. So far, though,
there has been no trace of such states in the LHC
data~\cite{aad2013search}.

\begin{figure}[!t]
\centering
\includegraphics[scale=0.22]{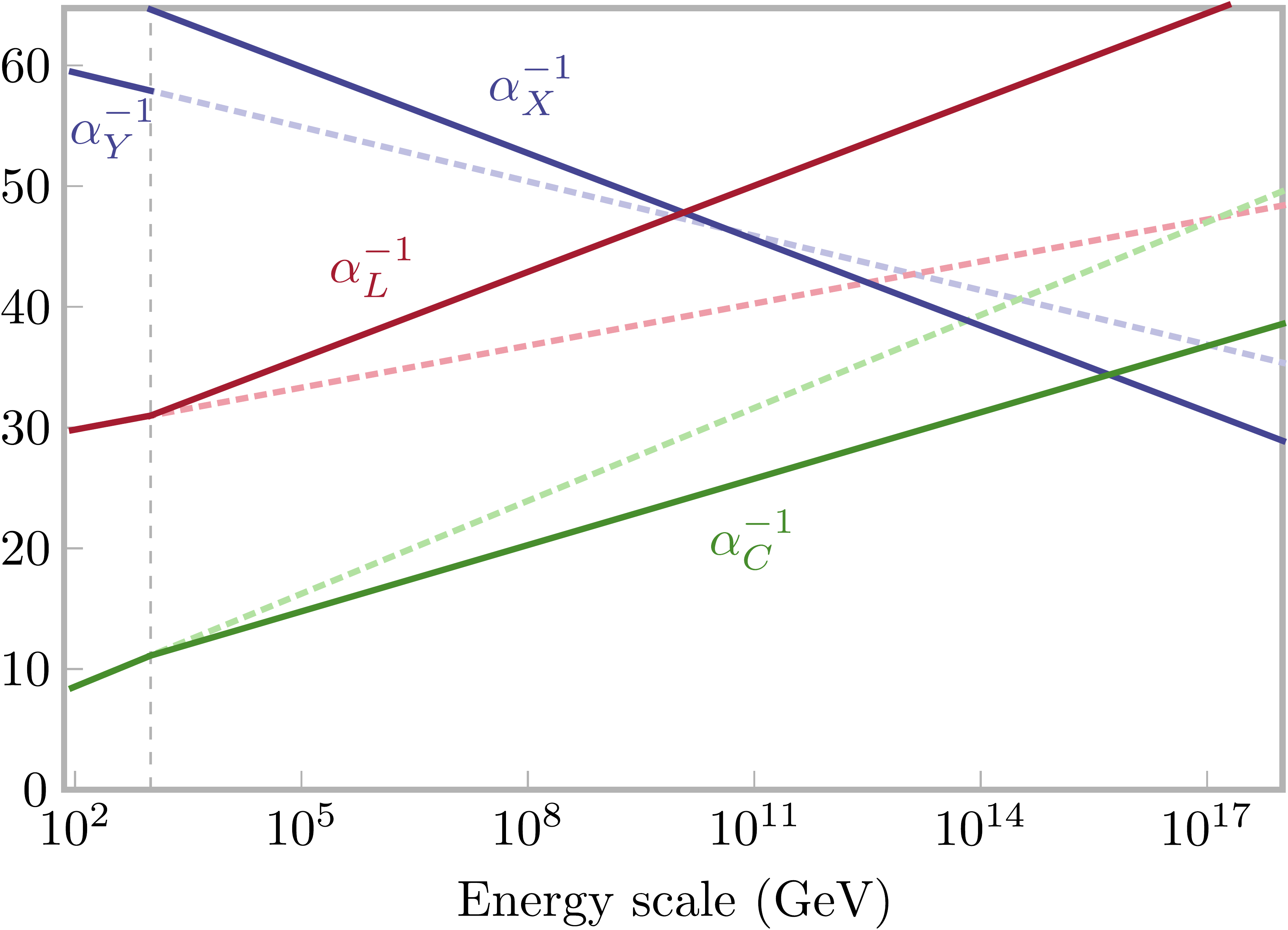} 
\caption{Running of the gauge couplings in the SM (dashed
  lines) and in the model in Ref.~\cite{PhysRevD.90.013005} (solid
  lines). Here the  $M_{331}$ scale is set to $1$~TeV. }
\label{fig:UnificationPlots0}
\end{figure}

Here we consider an alternative approach in which new physics at the
TeV scale realizes an extended electroweak gauge structure with
perturbatively conserved baryon-number. For definiteness we consider
the \TrTrOne~\cite{Singer:1980sw,valle:1983dk} (3-3-1) framework,
which implies that the number of generations equals the number of
colors, in order to cancel anomalies. This scheme has attracted
attention recently also in connection with B
physics~\cite{Buras:2014yna,Buras:2013dea,Buras:2012dp}, or flavor
symmetries~\cite{PhysRevD.81.053004,Vien:2014gza,Hernandez:2014lpa}.
In this letter we present a model in which the gauge couplings can
naturally unify at some accessible energy, and where small calculable
neutrino masses are induced by new gauge bosons exchange, in the
absence of supersymmetry.
Neutrino masses arise at the TeV scale~\cite{Boucenna:2014zba}
instead of the conventional high-scale seesaw
mechanism~\cite{PhysRevD.22.2227}.
We first recall that, by adding three gauge singlet fermions
$S_{i}^{}$, the light neutrinos acquire mass only at one-loop
order~\cite{PhysRevD.90.013005}.  Unfortunately, however, unification
does not occur, as can be seen in \fig{UnificationPlots0}. This is
mostly due to fact that the new gauge bosons make $\alpha_L$ weaker at
high energies, while the new colored particles strengthen $\alpha_C$.
Hence we contemplate the possibility of unifying the gauge couplings
in such a scheme~\footnote{For other RGE studies in the context of
  3-3-1 models see Ref.~\cite{Diaz:2005bw}.} by promoting the three
fermion singlets to three octets of the enlarged electroweak symmetry.
The new variant not only opens the possibility of reconciling neutrino
mass generation with gauge coupling unification but also provides a
novel radiative seesaw mechanism. The \TrOne~ gauge group is broken
down to the standard \TwoOne~ model at some scale $M_{331}$
characterizing the new gauge boson masses. This scale is found to lie
in the $1-10 \,\mathrm{TeV}$ range, with a plethora of new states
expected to be directly accessible to LHC searches.\\[-.8cm]

\section*{The model}

We consider a simple variant of the model introduced
in~\cite{PhysRevD.90.013005}, where the fermion singlet is now
promoted to an octet representation of $SU(3)_L$. The model is based
on the same \TrTrOne~ gauge symmetry, extended with a global
$U(1)_{\mathcal{L}}$, which is necessary in order to consistently
define lepton number, and an auxiliary parity symmetry whose purpose
will be made clear below.
 \begin{center}
  \begin{table}[b]
  \begin{centering}
  \begin{tabular}{cccccccccccc}
    \toprule 
    \multirow{2}{*}{} & \multicolumn{7}{c}{ Left-handed Fermions} &  & 
    \multicolumn{3}{c}{ Scalars}\tabularnewline \cmidrule{2-8} \cmidrule{10-12} 
    & $\psi_{L}$ & $\ell_{R}^{*}$ & $Q_{L}^{12}$ & $Q_{L}^{3}$ & $u_{R}^{*}$ & $d_{R}^{*}$ & $\Omega^{*}$ &  & 
    $\phi_{1}$ & $\phi_{2}$ & $\phi_{3}$\tabularnewline \midrule
    $SU(3)_{C}$ & $\boldsymbol{1}$ & $\boldsymbol{1}$ & $\boldsymbol{3}$ & $\boldsymbol{3}$ &  
    $\overline{\boldsymbol{3}}$ & $\overline{\boldsymbol{3}}$ & $\boldsymbol{1}$ &  & $\boldsymbol{1}$ & 
    $\boldsymbol{1}$ & $\boldsymbol{1}$\tabularnewline
    $SU(3)_{L}$ & $\overline{\boldsymbol{3}}$ & $\boldsymbol{1}$ & $\boldsymbol{3}$ &  
    $\overline{\boldsymbol{3}}$ & $\boldsymbol{1}$ & $\boldsymbol{1}$ & $\boldsymbol{8}$ &  & 
    $\overline{\boldsymbol{3}}$ & $\overline{\boldsymbol{3}}$ & $\overline{\boldsymbol{3}}$\tabularnewline
    $U(1)_{X}$ & $-\frac{1}{3}$ & $1$ & $0$ & $\frac{1}{3}$ & $-\frac{2}{3}$ & $\frac{1}{3}$ & $0$ &  & 
    $\frac{2}{3}$ & $-\frac{1}{3}$ & $-\frac{1}{3}$\tabularnewline \midrule
    $U(1)_{\mathcal{L}}$ & $-\frac{1}{3}$ & $1$ & $-\frac{2}{3}$ & $\frac{2}{3}$ & $0$ & $0$ & $0$ &  & 
    $\frac{2}{3}$ & $-\frac{1}{3}$ & $\frac{2}{3}$\tabularnewline
    $Z_{2}$ & $+$ & $-$ & $+$ & $-$ & $-$ & $+$ & $+$ &  & $-$ & $+$ & $+$\tabularnewline \midrule 
    multiplicity & 3 & 3 & 2 & 1 & 4 & 5 & 3 &  & 1 & 1 & 1\tabularnewline \bottomrule
 \end{tabular}
 \par\end{centering}
 \caption{\label{tab:field_content}Field content of the model.}
\end{table}
\par\end{center}
The model contains three generations of lepton $SU(3)_L$
anti-triplets, two generations of quark triplets and one of
anti-triplets (quarks and charged leptons are accompanied with their
right-handed $SU(3)_L$ singlet partners), three generations of fermion
octets, and finally three scalar boson anti-triplets. We summarize the
particle content of the model in~\tab{field_content}.
The allowed lepton interactions compatible with the quantum number
assignments given in~\tab{field_content} are the following:
 \begin{align}
 \label{eq:lag}
  \nonumber \mathscr{L}_{\textrm{leptons}} = &  
   \left( y_{ij}^{\ell} \right)^{*} \psi_{L,i}^{T} C \ell_{R,j}^{c} \phi_{1}^{*}    
  + \left( y_{ij}^{\prime} \right)^{*} \psi_{L,i}^{T} C \Omega_{j}^{c} \phi_{2}^{*} 
  \\ 
  & +\frac{1}{2}\left({\Mo}_{ij}\right)^{*}\left(\Omega_{i}^{c}\right)^{T}C\Omega_{j}^{c}+\textrm{h.c.}
 \end{align}

The components of $\psi_{L}$ and the $\phi_{j}$ are written as:
 \begin{equation}
  \psi_{L,i}=\left(\begin{array}{c}
   \ell_{L}\\
   -\nu_{L}\\
   N^{c}
   \end{array}\right)_i,\,
  \phi_{1}=\left(\begin{array}{c}
   \phi_{1}^{0}\\
   -\phi_{1}^{+}\\
   \widetilde{\phi}_{1}^{+}
   \end{array}\right),\,
  \phi_{2,3}=\left(\begin{array}{c}
   \phi_{2,3}^{-}\\
   -\phi_{2,3}^{0}\\
   \widetilde{\phi}_{2,3}^{0}
  \end{array}\right)\,.
 \end{equation}
 The scalars take vacuum expectation values (\textit{vevs}) in the
 directions $\vev{\phi_{1}}=(k_1,0,0)$ and
 $\vev{\phi_{2,3}}=(0,-k_{2,3},n_{2,3})$.  As for the octets, one can
 write
 \begin{align}
  \Omega_{i}^{c} & = 
  \left(\begin{array}{ccc}
   - \frac{1}{\sqrt{2}}T^{0} + \frac{1}{\sqrt{6}} \widetilde{N}^{c} & -T^{+} & \overline{\ell}_{L} \\
   - T^{-} & \frac{1}{\sqrt{2}} T^{0} + \frac{1}{\sqrt{6}} \widetilde{N}^{c} & -\overline{\nu}_{L} \\
   \widetilde{\ell}_{L} & -\widetilde{\nu}_{L} & -\frac{2}{\sqrt{6}}\widetilde{N}^{c}
  \end{array}\right)_{i}\,,
 \end{align}
 such that $\Omega_{i}^{c}$ is transformed into $U\Omega_{i}^{c}U^{\dagger}$
 under an $SU(3)_{L}$ gauge transformation, where $U$ is the
 transformation matrix of the triplet representation.\\

 Under \TwoOne, each $\Omega_{i}^{c}$ breaks into the representations
 $\left(\boldsymbol{3},0\right)\equiv\left(T^{+},T^{0},T^{-}\right)_{i}$,
 $\left(\boldsymbol{2},\frac{1}{2}\right)\equiv\left(\overline{\ell}_{L},-\overline{\nu}_{L}\right)_{i}$,
 $\left(\boldsymbol{2},-\frac{1}{2}\right)\equiv\left(\widetilde{\nu}_{L},\widetilde{\ell}_{L}\right)_{i}$,
 $\left(\boldsymbol{1},0\right)\equiv\widetilde{N}_{i}^{c}$, so there
 are four new charged leptons ($T^{+}$, $T^{-}$,
 $\widetilde{\ell}_{L}$, $\overline{\ell}_{L}$) and four new neutral
 fermion states ($T^{0}$, $\overline{\nu}_{L}$, $\widetilde{\nu}_{L}$,
 $\widetilde{\ell}_{L}$) in each generation.\\

 Note that the $U(1)_{\mathcal{L}}$ charge assignment of $\Omega_i$
 and $\phi_{2}$ differs from the one of the singlet of
 Ref.~\cite{PhysRevD.90.013005} so as to allow for a (vector-like)
 octet mass term, $\Mo$, in \eq{lag}~\footnote{This term is required
   in order to provide an adequately large mass to the new charged
   leptons.}.  On the other hand, the $Z_{2}$ symmetry forbids a
 $\psi_{L}\psi_{L}\phi_{1}$ coupling, which leads to the existence of
 one massless neutrino state in each generation, at
 the tree level.

\begin{figure}[t!]
\centering
\includegraphics[scale=0.22]{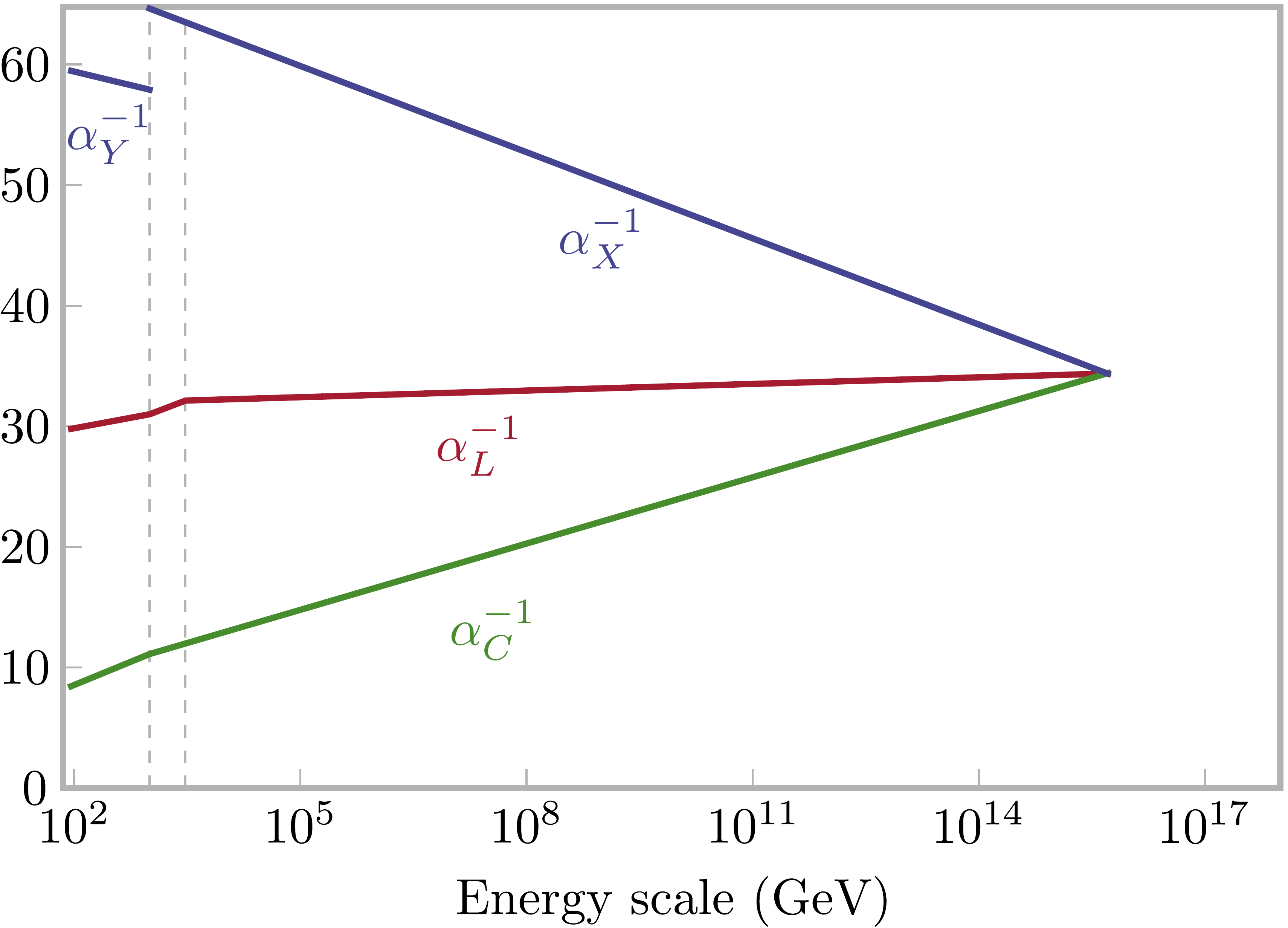}
\caption{Gauge coupling unification in the \TrTrOne~ model 
with three fermion octets with a $3$~TeV mass.}
\label{fig:UnificationPlots}
\end{figure}

 Note also that the electric charge and lepton number assignments of
 the particles of the model follow from~\footnote{As $T_8$ is a gauge
   generator, there is no physical Goldstone boson associated with
   spontaneous lepton number    violation~\cite{Chikashige1981265,PhysRevD.25.774}.}
\begin{eqnarray}
\label{eq:QL}
Q&=&T_3+\frac{1}{\sqrt{3}}T_8+X \,, \\
L&=&\frac{4}{\sqrt{3}} T_8+\mathcal{L} \,,
\end{eqnarray}
where $T_3$ and $T_8$ are the diagonal generators of $SU(3)_L$.\\[-.8cm]

\section*{Gauge coupling unification}

The one-loop renormalization group equation of the $\alpha_{i}\equiv
g_{i}^{2}/4\pi$ is given by~\cite{Gross:1973id,Politzer:1973fx}:
\begin{equation}
\frac{d\alpha_{i}^{-1}}{dt}=-\frac{b_{i}}{2\pi}\,,\label{eq:RGE}
\end{equation}
where $t$ is the logarithm of the energy scale, and the $b_{i}$ coefficients are functions of the Casimir of the
gauge group, $C\left(G_{i}\right)$, and of the Dynkin index of the scalar
and (Weyl) fermion representations, $T\left(S\right)$ and
$T\left(F\right)$, respectively:
\begin{equation}
b_{i}=-\frac{11}{3}C\left(G_{i}\right)+\frac{2}{3}\sum_{F}T\left(F\right)+\frac{1}{3}\sum_{S}T\left(S\right)\,.
\end{equation}

\begin{figure}[t!]
 \begin{centering}
  \includegraphics[scale=0.22]{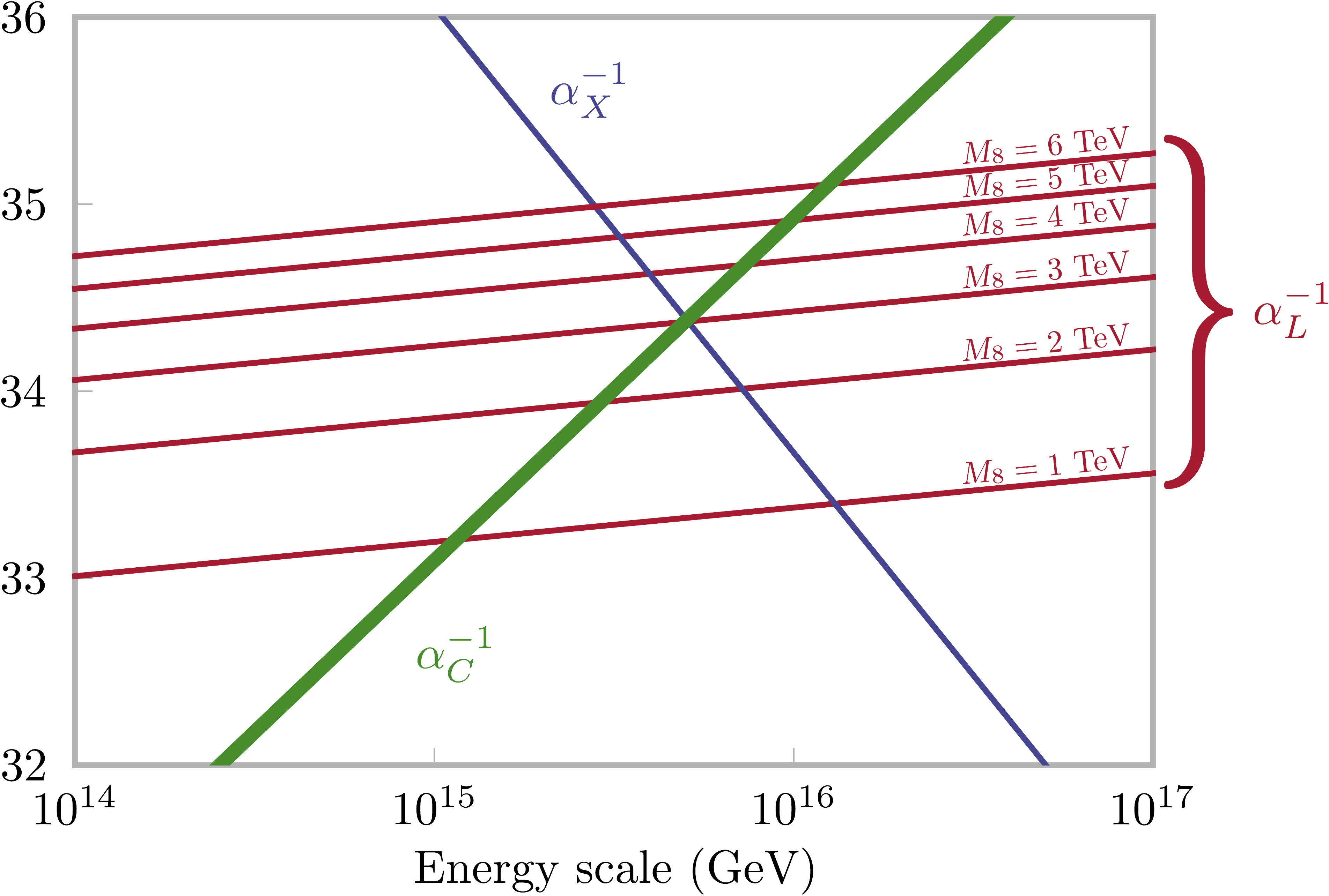}
  \par\end{centering}
\caption{\label{fig:UnificationPlot_Zoomed} Values of the gauge
  coupling constants near unification, assuming that the 3-3-1 scale
  is 1~TeV. The thickness of the $\alpha_{C}$ line reflects the
  $1\sigma$ uncertainty in the measurement of the strong coupling
  constant at the $m_{Z}$ scale~\citep{1674-1137-38-9-090001}, while
  the octet mass affects the running of $\alpha_{L}$. One sees how
  unification prefers $\Mo/M_{331}$ to lie roughly between 1 and 6.}
\end{figure}

For the SM, the $b_{i}$ are $b^\textrm{SM}=\left\{
  -7,-\frac{19}{6},\frac{41}{10}\right\} $, while in the
\TrTrOne~ phase they read
$b^\textrm{331}=\left\{ -5,-\frac{13}{2}+2n,\frac{13}{2}\right\} $,
for $n$ active fermion octets $\Omega_i$. It should be noted that
while we do not speculate here about the possible embedding of
\TrTrOne~ into some bigger group (for
example $E_{6}$), it can be shown on very general grounds that the
$U(1)_{X}$ charge normalization
should be $X_{\textrm{canonical}}=\sqrt{3}/2X$. Given the relation
between $X$ and the SM hypercharge indicated by \eq{QL},
it follows that $\alpha_{Y}^{-1}=\frac{1}{5}\left(\alpha_{L}^{-1}
+4\alpha_{X}^{-1}\right)$ at the 3-3-1 breaking scale.
Figure~\ref{fig:UnificationPlots} illustrates the running of the gauge
coupling constants in our model, with the 3-3-1 scale fixed at 1 TeV
and the three octets $\Omega_i$ integrated out at 3 TeV.
The exotic scalar states are also integrated out at the 1 TeV scale,
although the running of the gauge couplings is not very sensitive to
this value.  Given that the $b_{i}^\textrm{SM}$ coefficients are not
very different from $b_{i}^\textrm{331}$ with the three octets,
unification is sensitive mostly to the ratio $\Mo/M_{331}$, and not to
the 3-3-1 scale \textit{per se}.
The effect is shown in \fig{UnificationPlot_Zoomed}; allowing for
threshold and 2-loop effects, one can see that unification constrains
$\beta^{-1}\equiv M_8/M_{331}$ to lie between 1 and 6.\\[-.8cm]

\section*{Mass matrices}

Lepton mass matrices arise from the Lagrangian in \eq{lag} after
\TrOne~ breaks to $U(1)_{\textrm{em}}$,
 \begin{equation}  
 \mathscr{L}_{\textrm{lepton masses}}^{\ell,\nu} =   
   \boldsymbol{\ell}^{T}C \mathcal{M}_{\ell}^{*} \boldsymbol{\ell^{c}} 
   + \frac{1}{2} \boldsymbol{\nu}^{T} C \mathcal{M}_{\nu}^{*} \boldsymbol{\nu} + \textrm{h.c.} 
 \end{equation}
 In the basis where $\boldsymbol{\ell}=\left(\ell_{L},
   \widetilde{\ell}_{L}, T^{-}\right)^T$ and
 $\boldsymbol{\ell^{c}}=\left(\ell_{R}^{c},\overline{\ell}_{L},T^{+}\right)^T$,
 the charged leptons mass matrix reads
\begin{equation}
 \mathcal{M}_{\ell}=
  \begin{pmatrix} 
   M^{\ell} & M_{331}^\prime & M_W^\prime \\
   0 & \Mo & 0 \\
   0 & 0 & \Mo
  \end{pmatrix}\,,
\end{equation}
where the entries are given by
\begin{equation}
(M^{\ell})_{ij} \equiv y_{ij}^{\ell}k_{1},\, (M_W^\prime)_{ij}\equiv y_{ij}^{\prime}k_2, 
   \textrm{ and }  (M_{331}^\prime)_{ij} \equiv y_{ij}^{\prime}n_2\,.
   \nonumber
\end{equation}

Note that the \textit{vev} $n_2$ (together with $n_3$) sets the
\TrOne~ breaking scale.  In contrast, the \textit{vevs} $k_{1}$ and
$k_2$ must lie at the electroweak scale since they belong to $SU(2)_L$
doublets after the breaking of $SU(3)_L$.\\

From these expressions it follows that amongst the charged leptons,
there is only a pair which is light in each generation, namely:
 \begin{align}
  \ell_{\textrm{light}} & \propto \ell_{L} - x \widetilde{\ell}_{L} - x\alpha T^{-}\,,\\
  \ell_{\textrm{light}}^{c} & \propto\ell_{R}^{c} - \frac{ M^{\ell} }{ M_{331}^\prime }
   \frac{ x^{2} }{ x^{2} + 1 } \overline{\ell}_{L}\,,
 \end{align}
 where $\alpha\equiv M_W^\prime/M_{331}^\prime=k_2/n_2$ and $x\equiv
 M_{331}^\prime/\Mo$.
 Here the parameter $x=y'\beta$ is constrained by unification to lie
 in the range $y'/6\lesssim x \lesssim y'$.
 These two 2-component states form the standard Dirac charged lepton
 which now has a squared mass given by $\left(M^{\ell}\right)^{2}
 \frac{ 1 }{ 1 + x^{2} }$, an expression that differs from the SM one.
 Notice that the presence of states which do not come from the
 $\left(\boldsymbol{2},-\frac{1}{2}\right)$ electroweak representation
 is $\alpha$-suppressed. The two remaining pairs of charged leptons
 are heavy, with octet-scale masses.\\

 Turning now to the neutral fermions,  their mass matrix
 reads: \\

\noindent \resizebox{\linewidth}{!}{%
$\mathcal{M}_{\nu} =
  \begin{pmatrix}
   0 & 0 & 0 & M_{331}^\prime & \frac{1}{ \sqrt{6} } M_W^\prime & \frac{1}{ \sqrt{2} } M_W^\prime \\
   0 & 0 & {M_W^\prime} & 0 & -\frac{2}{\sqrt{6}}M_{331}^\prime & 0 \\
   0 & {M_W^\prime}^T & 0 & \Mo & 0 & 0 \\
   {M_{331}^\prime}^T & 0 & \Mo & 0 &0 & 0 \\
   \frac{1}{\sqrt{6}}{M_W^\prime}^T &  -\frac{2}{\sqrt{6}}{M_{331}^\prime}^T & 0 & 0 & \Mo & 0\\
   \frac{1}{\sqrt{2}}{M_W^\prime}^T & 0 & 0 & 0 & 0 & \Mo
  \end{pmatrix}\,,$
}\\
in the eigenbasis
$\boldsymbol{\nu}=\left(\nu_{L},N^{c},\widetilde{\nu}_{L},
  \overline{\nu}_{L},\widetilde{N}^{c},T^{0}\right)^T$.
In the one-family approximation, $\mathcal{M}_{\nu}$ has a null eigenvector
\begin{equation}
\nu_{\textrm{light}}=\sum_{\alpha}\omega_{\alpha}\nu_{\alpha}\,,
\end{equation}
with $\omega$ given as $\frac{1}{N} \left( 1, -\alpha,-x, x
  \alpha^{2}, -\sqrt{ \frac{3}{2} } x \alpha, - \frac{1}{\sqrt{2}} x
  \alpha \right)^T$, where $N$ is some normalization factor.
Note that the observed neutrinos are mainly a mixture of $\nu_{L}$ and
$\widetilde{\nu}_{L}$ which are both in the
$\left(\boldsymbol{2},-\frac{1}{2}\right)$ representation of the
\TwoOne~ group. The admixture of the remaining neutrino states are
suppressed by at least a factor $\alpha$, which can be vanishingly
small.
We have verified that in the multi-generation case $\mathcal{M}_{\nu}$
has a null eigenvector associated to each of the three generations of
leptons. As a result neutrinos are massless in the tree level
approximation. This property forms the basis of the radiative
mechanism discussed below.\\
\begin{figure}[!t]
 \begin{centering}
  \includegraphics[scale=0.75]{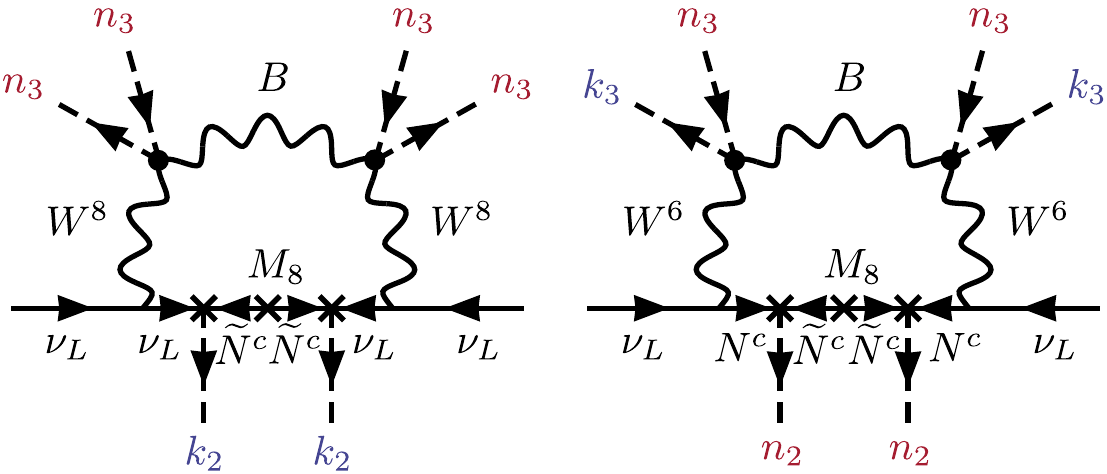}
  \par\end{centering}
\caption{\label{fig:MassInserionExample} Diagrams contributing to
  light neutrino mass.}
\end{figure}

At the one-loop level, the exchange of gauge bosons will give rise to
a dimension-nine operator which, after symmetry breaking, yields a
small neutrino mass through diagrams such as those displayed in
Fig.~\ref{fig:MassInserionExample}.
In order to understand the result of a detailed exact calculation here
we simply focus on a typical contribution illustrated in
Fig.~\ref{fig:MassInserionExample}, which is found to be of the form
 \begin{align}\label{eq:numass}
  m_{\nu} & \approx \frac{1}{ \left( 4 \pi \right)^{2} } \frac{1}{36} g_{L}^{4} 
  \left( 3g_{L}^{2} + 4g_{X}^{2} \right) n_3^{2} 
  \left( k_3 n_2 - k_2 n_3 \right)^{2} \Mo {y^\prime}^{2} \nonumber \\
  & \times f \left(\left\langle m_{A}^{2}\right\rangle ,
\left\langle m_{\Psi}^{2}\right\rangle \right)\,,
 \end{align}
 where $f$ is some loop function with dimensions of
 $\left(\textrm{mass}\right)^{-6}$, depending on the internal masses.
 This approximation captures the key features of the exact result such
 as ($i$) the gauge nature of the underlying radiative dimension-nine
 seesaw mechanism, requiring two bosonic and three fermionic mass
 insertions in the internal lines; as well as ($ii$) the symmetry
 structure, both gauge as well as lepton number, as can be seen
 explicitly. One also sees that no mass is generated in the limit
 where $k_3 n_2 - k_2 n_3$ is set to zero~\footnote{As usual in mass
   insertion methods, they are not accurate enough for reliable
   estimates. In our case this is manifest as uncertainties in the
   choice of the ``average'' mass parameters $\vev{ m_{A}^{2}},\vev{
     m_{\Psi}^{2}}$.}.
 With a \textit{vev} alignment which minimizes this factor, neutrinos
 get a sub-eV mass even for large $y^\prime$ values.
 Should $y'$ be significantly smaller than unity, one must suppress
 the mixing between $N^{c}$ and $\nu_L$, since in this case one of the
 neutrino mass eigenstates is mainly $N^{c}$, with approximate mass
 $\frac{2}{3}\left(M_{331}^{\prime}\right)^{2}/\Mo$. This is readily
 achieved by setting $\alpha \to 0$. \\[-.8cm]

\section*{Summary and outlook}

In this letter we have proposed an electroweak \TrTrOne~ gauge
extension of the SM in which neutrinos are massless at
tree-level. Even though the neutral fermion mass matrix has a seesaw
structure, the messengers only provide mass at the loop level, thanks
to the symmetry protection. Gauge mediated radiative corrections
generate small calculable neutrino masses.
The physics responsible for providing small neutrino masses is also
responsible for gauge coupling unification, which can be achieved at a
characteristic scale of order TeV in the absence of supersymmetry and
of GUT-like interactions. A plethora of new states such as new gauge
bosons and fermions makes the model directly testable at the LHC, with
a non-trivial interplay between the quark sector and the lepton
sector. The presence of such new features is presently under
investigation.\\

We thank Martin Hirsch for valuable discussions. This work was
supported by Spanish grants FPA2011-22975 and Multidark CSD2009-00064
(\textit{MINECO}), and PROMETEOII/2014/084 (\textit{Generalitat
  Valenciana}). FGC acknowledges support from \textit{CONACYT} under
grant 208055. RF was supported by 
\textit{Funda\c{c}\~ao para a Ci\^encia e a Tecnologia} through grants
CERN/FP/123580/2011 and EXPL/FIS-NUC/0460/2013.


%

\end{document}